\documentclass[%
 reprint,
 amsmath,amssymb,
 aps,
]{revtex4-2}

\usepackage{graphicx}
\usepackage{dcolumn}
\usepackage{bm}
\usepackage[utf8]{inputenc} 
\usepackage[T1]{fontenc}    
\usepackage{hyperref}       
\usepackage{url}            
\usepackage{booktabs}       
\usepackage{amsfonts}       
\usepackage{nicefrac}       
\usepackage{microtype}      
\usepackage{lipsum}
\usepackage{graphicx}
\usepackage{amssymb,amsthm,amsmath}
\usepackage{tabularx}
\usepackage{graphicx,subcaption,lipsum}
\usepackage{float} 
\usepackage{xcolor,paralist,hyperref,titlesec,fancyhdr,etoolbox}
\usepackage{epigraph}

\begin{document}

\preprint{APS/123-QED}

\title{Climbing the ladder: a method for identifying promising copper-lead apatites}

\author{Ihor Sukhenko}
 \email{isukhenko@imp.kiev.ua}
\author{Volodymyr Karbivskyy}%
\affiliation{%
 Institute of Metal Physics, NAS of Ukraine\\
 36 Vernadsky blvd, 03142 Kyiv, Ukraine
}%




\date{\today}

\begin{abstract}
We develop a DFT screening procedure for copper-substituted lead apatites of the composition Pb$_9$Cu(XO$_4$)$_6$Y that enforces three design rules: thermodynamic stability, Cu site preference, and symmetry robustness of the near-Fermi electronic structure. A convex-hull analysis over P/V/As as X, O/F/Cl/Br as Y, identifies vanadates as the only members on or beneath the hull. Across the family, Cu substitution at Pb$^{\text{I}}$ ($4f$) preserves flat bands at E$_F$, whereas Pb$^{\text{I}}$ ($6h$) either gaps or severely distorts them. Small symmetry-lowering relaxations ($P3 \rightarrow P1$) are also capable of opening the band gap, motivating symmetry robustness as a filter. Applying these criteria singles out Pb$_9$Cu(VO$_4$)$_6$Br$_2$ (and, possibly, Cl$_2$) as leading candidates. This work motivates experimental study of the selected compounds, as well as a dedicated study of strong correlations. 
\end{abstract}

\maketitle


\section{\label{sec:intro} Introduction}

Copper-substituted lead apatites have attracted the attention of the community due to early claims of the discovery of high-temperature superconductivity (HTS) that later turned out not to be substantiated \cite{Sukhenko2024}.

Despite this, closer theoretical analysis of the system under focus revealed potentially worthwhile electronic structure. Already the particular claimant compound («LK-99») was a strongly correlated material \cite{mott-or-charge, correlated} with, supposedly, a couple of flat bands at or near the Fermi level featuring two Weyl points, possessed non-trivial quantum geometry \cite{Weyl} and magnetism \cite{Jiang}. The interest in this family of compounds has waned as a result of the refutation of the HTS claim. However, it is conceivable that, for this reason, interesting physics may be overlooked. In this paper, we pose the question: What would it take for a member of the crystalline family Pb$_9$Cu(XO$_4$)$_6$Y to preserve the electronic-structure portrait described above? To answer this question, we first briefly revisit the literature on now-infamous LK-99 to discern and highlight the causes that prevented the theorised electronic structure from manifesting in reality in its case. We also stick to one-particle paradigm and do not analyse superconductivity itself. Detailed treatment of the correlation effects is reserved for another study.

The first issue we encounter is the thermodynamic stability. Shen et al. \cite{phase_stability} showed that Pb$_9$Cu(PO$_4$)$_6$O was either unstable or metastable, in danger of decomposition into other phases.

Another sticking point was the site preference. The apatite structure has two inequivalent crystallographic positions - Pb$^{\text{I}}$ (Wyckoff $4f$) and Pb$^{\text{II}}$ ($6h$) which can host a substituent atom. It turns out that the resulting band structure is substantially different between the two scenarios \cite{iowa, sinead}, with only one of them leading to the expected band structure.

Lastly, it was predicted that the breaking of symmetry all the way to triclinic inflicted by the copper substitution and subsequent lattice destabilisation might also open the gap turning the compound into a semiconductor \cite{iowa, chile-triclinic} (as, incidentally, was observed in experiment \cite{monocrystal, ferrom_china, china_semicond, india_pure, taiwan_cu2s}).

Instead of being unique to LK-99 replication studies, which are virtually a closed chapter, all these issues are highly likely to persist whenever one attempts to study similar compounds of the same crystal family.

Here we propose a method that would allow us to select promising candidate compounds among possible copper-substituted lead apatites while simultaneously addressing all those issues. 

The goal is to realise that «flat bands at the Fermi level» picture that sparked the initial interest. The chosen candidate compounds will then be robust against those three filters: 

\begin{itemize}
    \item Thermodynamic stability,
    \item Copper site preference,
    \item Symmetry preservation.
\end{itemize}

The set of compounds that underwent our screening procedure consists of: Pb$_9$Cu(XO$_4$)$_6$Y$_2$, where X = P, V, As; Y = O$_{1/2}$, F, Cl, Br. The choice of X and Y elements is natural as far as the apatite family goes, as these ones are the most common in nature, applications and the literature  \cite{karb, prediction-ml, White2003, White2005, Kim2005, Ardanova2010, kanazawa}.

After selecting the most suitable candidate compounds we shall further analyse them in some detail, and consider their magnetic ground state and the band structure.

\section{\label{sec:level2} Theory and methods}

\subsection*{Thermodynamic stability}

The convex hull method is used to predict an approximate estimate of the thermodynamic stability of compounds, in particular those that have not yet been obtained experimentally. For a candidate of composition vector $\vec{\phi}$ (e.g. $\vec{\phi} = \left( \phi_{\text{Pb}}, \phi_{\text{Cu}}, \phi_{\text{P}}, \phi_{\text{O}} \right)$ ...) (with $\sum_i \phi_i=1$), the per-atom formation enthalpy is
\begin{equation}
\Delta H_f = E_{\rm DFT} - \sum_i \phi_i E^{\text{I}}_{\rm DFT},
\end{equation}
where \(E_{\rm DFT}\) is the total energy of the compound and
\(E^{\text{I}}_{\rm DFT}\) are the elemental reference energies at 0 K.

Although negative $\Delta H_f$ is a necessary condition for the stability of the compound, it is not sufficient, because competing phases may be more favourable. Therefore, the indicator of stability in this approach is the \textit{distance from the hull} $\Delta \varepsilon_{hull}$. It is determined as follows:

\begin{equation}\label{eq:ehull}
\Delta E_{\rm hull}(\boldsymbol{\phi})
= \Delta H_f(\boldsymbol{\phi}) -
\min_{\{\lambda_j\}} \sum_j \lambda_j \Delta H_f(\boldsymbol{\phi}_j),
\end{equation}

subject to \(\lambda_j \geq 0\), \(\sum_j \lambda_j = 1\), and
\(\sum_j \lambda_j \boldsymbol{\phi}_j = \boldsymbol{\phi}\).
This minimisation mixes known competing phases \(\{\boldsymbol{\phi}_j\}\)
and is solved by a linear program. Compounds on the hull satisfy
\(\Delta E_{\rm hull} \le 0\); otherwise \(\Delta E_{\rm hull} > 0\)
is the decomposition driving force, and the minimising
\(\{\lambda_j\}\) specify the decomposition products.

It is, however, important to remember, that the convex-hull stability is a 0 K criterion that neglects finite-temperature effects such as vibrational and configurational entropies. Nevertheless, it is suitable for an initial assessment of the suitability for synthesis when discovering new materials \cite{shen}.

\textit{Data and implementation}. It is obvious from the construction of the method that the more points the convex hull contains in a given compositional space, the more accurate the interpolation will be, and the more reliable the result. Materials databases created on the basis of high-performance DFT calculations come to the rescue. Our choice was the Open Quantum Materials Database (OQMD) \cite{oqmd}, which at the time of writing contains 1,226,781 materials. Practically, the whole procedure is as follows. The compositional space in which the desired compound is located is unloaded from OQMD. Then only compounds with the lowest energy for a given composition are selected. The energy $\Delta H_{hull}(\vec{\phi})$ is calculated within a linear programming algorithm described by \ref{eq:ehull}.

\subsection*{Symmetry breaking and metallicity}

\epigraph{The focus on quantum materials has raised questions on the fitness of density functional theory for the description of the basic physics of such strongly correlated systems. Recent studies point to another possibility: the perceived limitations are often not a failure of the density functional theory per se, but rather a failure to break symmetry.}{\textit{Alex Zunger} \cite{zunger}}

It has become almost a cliché to say that the density functional theory fails at strong correlations, supposedly due to its mean-field appearance that precludes it from handling two-particle processes. Recently, however, cracks have started to appear in this view. In \cite{cuprate-scan} by using the SCAN functional \cite{scan} and by breaking the symmetry of the system, a correct doping behaviour, metal-to insulator transition (MIT) and antiferromagnetism of La$_2$CuO$_4$, a parent compound to cuprate superconductors, was reproduced. Stripe and magnetic order in the famous YBCO, as well as its MIT were properly addressed in \cite{ybco-dft} - again, by using SCAN and breaking the symmetry.

Why the breaking of symmetry is important?

For any finite isolated system, the exact ground state preserves all symmetries of the Hamiltonian. In macroscopic solids, however, expectation values of non-commuting observables can «freeze» as system size grows, resulting in an observable symmetry breaking even though an exact symmetry-unbroken wavefunction still exists in principle. Approximate density functionals can often select such broken-symmetry solutions, that in fact mimic the long-lived collective modes.

Strong correlation arises when several Slater determinants of the same symmetry are nearly degenerate and strongly mixed. Standard approximate functionals cannot represent that multi-determinant mixing; they tend to overestimate the energy of the symmetry-unbroken state. Allowing symmetry breaking lifts the near-degeneracy and replaces «true» correlation with a mean-field order whose energy these functionals can describe accurately-thus recovering part of the missing correlation energy by imitation \cite{perdew-symmetry-carbon}. An elegant in-depth perspective on this problem can be found in \cite{perdew-interpretation}.

For our task, the symmetry check is performed as follows. Atomic coordinates and lattice parameters are relaxed while 1) fixing the unbroken $P3$ lattice symmetry;
2) under the broken $P1$ symmetry, accompanied by the addition of random values of the order of $10^{-5}$ \AA{} to the crystal coordinates, $10^{-3}$ \AA{} to the lattice parameters, and $10^{-3 \circ}$ to the angles to push the structure away from the local energy minimum. The resulting total energy values for both symmetries are compared, and band structures are calculated.

\subsection*{Magnetism}

\begin{figure}
	\centering
	\includegraphics[width=0.9\linewidth]{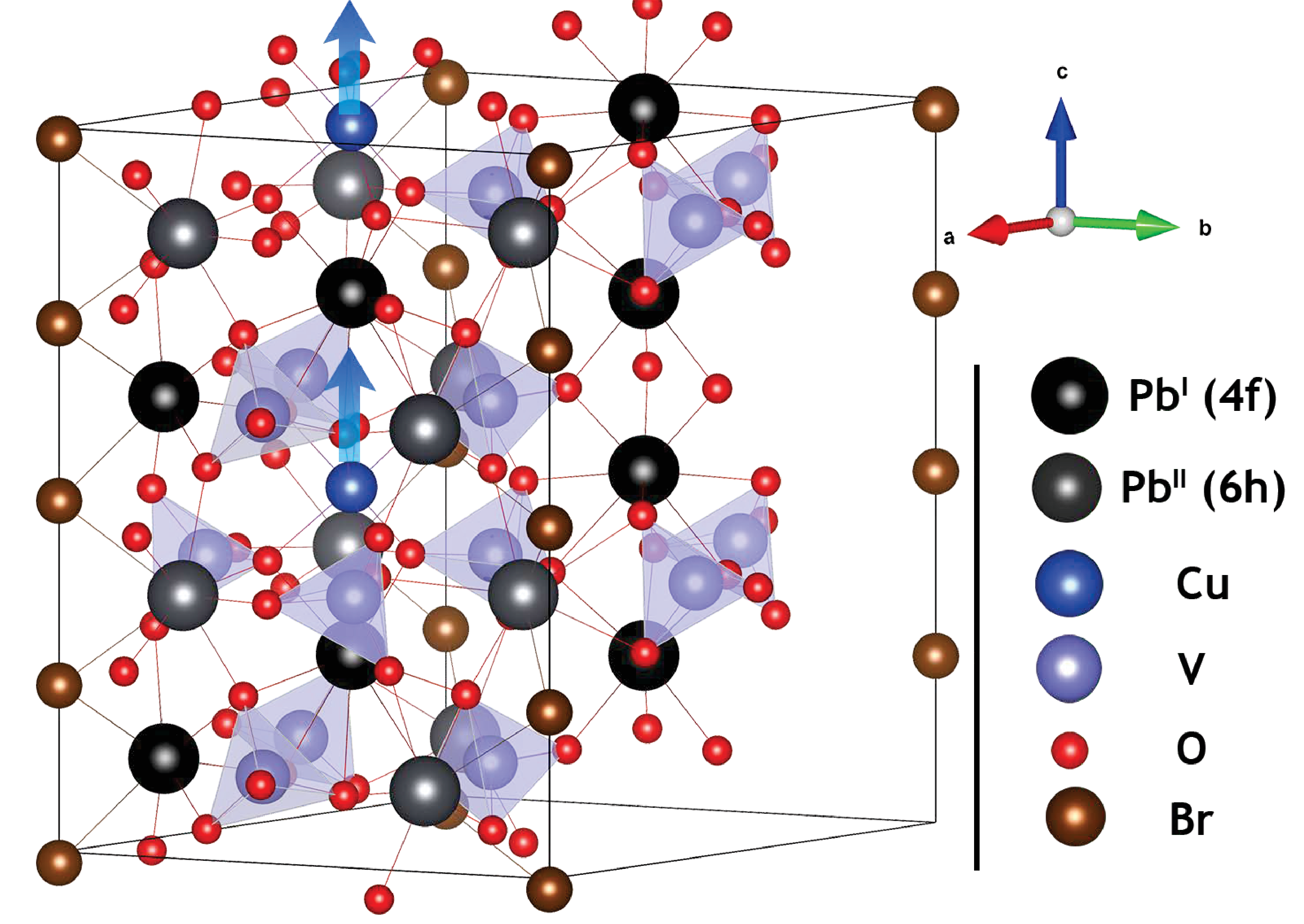} 
	\caption{Illustration of a Pb$_9$Cu(VO$_4$)$_6$Br$_2$ supercell with ferromagnetic spin ordering.}
	\label{fig:fm-scheme}
\end{figure}

The Cu $3d^9$ configuration (nominal $S = 1/2$) motivates an assessment of magnetic interactions and possible long-range order.
A common route is the energy mapping: compute total energies for many spin configurations (often in large supercells) and fit a Heisenberg model. While straightforward, this becomes costly, especially considering the fact that our compounds of interest have 41-42 atoms in the unit cell.

We instead extract exchange parameters from a \emph{single} primitive-cell calculation using the Green’s-function (magnetic force theorem) formalism as implemented in \textsc{TB2J}~\cite{tb2j}. Within the Liechtenstein-Katsnelson-Antropov-Gubanov (LKAG) \cite{lkag} framework, a small rigid rotation of local spins is treated as a perturbation to the one-electron Hamiltonian (in practice, of Wannier form). The resulting energy change maps analytically onto an extended Heisenberg model,
\begin{equation}
\mathcal{H} = -\sum_{i\neq j} J_{ij} \vec{S_i} \cdot \vec{S_j}
- \sum_{i\neq j} \mathbf D_{ij}\cdot (\vec{S_i} \times \vec{S_j})
- \sum_{i\neq j} \vec{S_i} \mathbf{J}_{ij}^{ani} \vec{S_j} ,
\end{equation}
where (by TB2J convention) $J_{ij}>0$ favours ferromagnetic alignment. In the collinear, non-relativistic case the isotropic exchange follows the LKAG expression is
\begin{equation}
J_{ij} = -\frac{1}{4\pi}\int_{-\infty}^{E_F} 
\text{Im } \text{Tr} \left[ \Delta_i G^{\uparrow}_{ij}(\varepsilon) \Delta_j G^{\downarrow}_{ji}(\varepsilon) \right] d\varepsilon,
\end{equation}
with spin splittings $\Delta_{\text{I}}$ and intersite single-particle Green's functions $G^{\sigma}_{ij}$. With spin–orbit coupling, \textsc{TB2J} also returns the Dzyaloshinskii–Moriya vectors $\mathbf D_{ij}$ and the symmetric anisotropic exchange $\mathbf{J}_{ij}^{ani}$ via the corresponding Green’s function expressions. For our purposes, DMI and anisotropic components were not considered.

Using WANNIER90 \cite{wannier90}, we construct a Wannier tight-binding Hamiltonian including the magnetic $d$ orbitals and ligand O-$p$, halogen-$p$ states, verify the frozen/disentanglement windows, and evaluate $J_{ij}$ up to a real-space cutoff $r_{\mathrm{cut}}$.

Compared to supercell energy mapping, the Green’s-function route (a) avoids enumerating many magnetic patterns, (b) is naturally long-ranged, (c) provides orbital-resolved decompositions, and (d) if needed, delivers DMI/anisotropy from the same primitive-cell input.







\subsection*{Computational details}

For DFT calculation, Quantum ESPRESSO package \cite{QE-2009,QE-2017,QE-2020} was used, in conjunction with Meta-GGA r$^2$SCAN (regularised restored strongly constrained and appropriately normed functional \cite{r2scan}) and norm-conserving pseudopotentials \cite{oncv}. For structural optimisations, A 4×4×6 k-point mesh was used for structural optimisations, and a 8×8×10 mesh - for band structure and density of states calculations. The plane-wave energy cutoff was set at 953 eV (70 Ry). For the elucidation of the band structure and the magnetic order, DFT+U+J was used, with U and J parameters determined from cRPA calcultaions performed with RESPACK \cite{respack, wan2respack} together with WANNIER90 \cite{wannier90}. In particular, the interaction parameters for Cu site of Pb$_9$Cu(VO$_4$)$_6$Br$_2$ were obtained as $U = 5.07$ eV, $J = 0.46$ eV.

\section{Results and Discussion}

\subsection*{Thermodynamic stability}

 Table \ref{tab:enthalpy} shows the values of the formation enthalpy per atom for the considered compounds; table \ref{tab:chemstab} shows the values of the shell distances for the corresponding composition space.

\begin{table}[b]
	\caption{Formation enthalpies, eV/unit cell}
	\begin{ruledtabular}
        \begin{tabular}{ccccc}
		& O & F & Cl & Br \\ 
            \hline
		P & -1.802 & -1.859 & -1.769 & -1.753 \\ 
		V & -2.349 & -2.410 & -2.332 & -2.309 \\ 
		As & -1.164 & -1.243 & -1.164 & -1.158 \\ 
	\end{tabular}
        \end{ruledtabular}
	\label{tab:enthalpy}
\end{table}

\begin{table}[b]
	\caption{Distance to the hull, eV}
        \begin{ruledtabular}
	\begin{tabular}{ccccc}
		 X$\textbackslash$Y & O & F & Cl & Br \\ 
         \hline
		P & 0.333 & 0.338 & 0.345 & 0.329 \\
		V & -0.416 & -0.388 & -0.384 & -0.398 \\
		As & 0.357 & 0.357 & 0.375 & 0.338 \\ 
	\end{tabular}
        \end{ruledtabular}
	\label{tab:chemstab}
\end{table}

It was found that all the considered compounds have a negative formation enthalpy. However, \textit{only vanadates} exhibit a negative distance to the hull. The fact that most other compounds do not can be understood by taking into account the difference in ionic radius and electronegativity between copper and lead, which destabilises the structure. But the uniqueness of vanadates, somewhat counter-intuitively, may be associated with the «softness» of VO$_4$ tetrahedra compared to AsO$_4$ and PO$_4$. Vanadium contributes to VO$_4$'s chemical bond of by $d$-orbitals, due to which the bond is weaker and more anisotropic. The latter fact is confirmed, in particular, by the significantly lower decomposition temperature of vanadium apatites experimentally \cite{vo4-thermal-1, vo4-thermal-2}. So, if for phosphates and arsenates it turns out to be possible to «disassemble» the apatite structure and build other compounds from the «bricks» AsO$_4$ and PO$_4$, such as Pb$_3$(PO$_4$)$_2$, Cu$_3$(PO$_4$)$_2$, for vanadates such an imaginary operation turns out to be impossible without destroying the «bricks» VO$_4$ themselves, which requires additional energy. Because of this, there are no oxides on the hull $\{$ Pb-Cu-V-O-Y $\}$ into which the considered samples would decompose. From a structural point of view, VO$_4$, being more flexible and slightly more voluminous, can somewhat alleviate the stress created when replacing lead with copper in the 4$f$ position.


\subsection*{Site preference, its role and origin}

\begin{figure}
        \centering
        \includegraphics[width=1.0\linewidth]{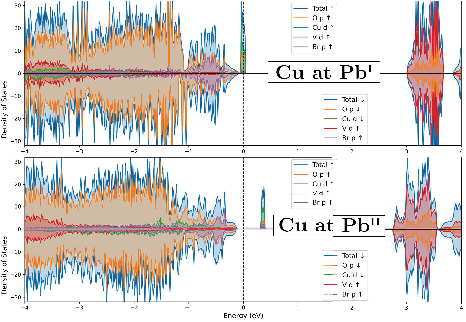} 
        \caption{Full and partial density of states of the compound Pb$_9$Cu(VO$_4$)$_6$Br$_2$ with two scenarios of copper substitution. Qualitatively, the same picture is observed for all considered compounds.}
        \label{fig:pb1vspb2}
\end{figure}

The choice of crystallographic position of the copper atom has a decisive influence on the electronic structure. Thus, our calculations show that for all compounds considered, substitution with copper at the Pb$^{\text{I}}$ position ($4f$ Wyckoff, Fig. \ref{fig:fm-scheme}), in a state $3d^{9} (Cu^{2+})$, leads to the desired form of the band structure. In contrast, substitution at the Pb$^{\text{II}}$ position ($6h$) opens a gap, turning the compound into an insulator (Fig. \ref{fig:pb1vspb2}).

The table \ref{tab:position} shows the difference in total energies per atom between the two substitution scenarios $|E(CuI)| - |E(CuII)|$.

\begin{table}[b]
	\caption{Energy difference between scenarios Cu$\rightarrow$PbI Cu$\rightarrow$PbII, meV/atom}
        \begin{ruledtabular}
	\begin{tabular}{ccccc}
		X\textbackslash{}Y & O & F & Cl & Br \\ 
            \hline
		P & 3.84 & 4.49 & 26.68 & 21.51 \\ \
		V & 19.06 & 11.34 & 19.35 & 13.72 \\ 
		As & 22.37 & 13.91 & 7.04 & 20.52 \\ 
	\end{tabular}
        \end{ruledtabular}
	\label{tab:position}
\end{table}

	

In the non-magnetic case without the spin-orbit coupling, Cu$\rightarrow$Pb$^{\text{I}}$ scenario preserves a global $C_3$ axis (space group $P3$). At $\Gamma$ and $A$ points, time-reversal with $\mathcal{T}=+1$ symmetry brings together the complex $C_3$ eigenvalues $e^{\pm i 2 \pi /3}$, forcing a band touching at these points. By contrast,Cu$\rightarrow$Pb$^{\text{II}}$ lowers the global symmetry and removes the $C_3$ axis; with only mirror symmetry left, no complex representation exists, and therefore no enforced twofold touching. Any apparent crossing is accidental, but when the spin polarisation is turned on, we observe the opening of the gap in all studied cases.

These results agree broadly with the findings of \cite{iowa} which found the same picture in application to Pb$9$Cu(PO$_4$)$_6$O compound.

Fortunately, our calculations show a predominant occurrence of copper in the first position in all the cases considered. However, the difference per atom is often less than $kT$ at room temperature. This fact prompts experimental determination of the distribution of these substitutions. Still, for further modelling, it is satisfactory to assume the Pb$^{\text{I}}$ position as the predominant one.

\subsection*{Symmetry}

\begin{table}[b]
	\caption{Total energy of the triclinic and trigonal phases (eV per unit cell), |E(P1)| -|E(P3)| for the series of compounds Pb$_9$Cu(YO$_4$)$_6$Z, where Y = P, V, As; Z = O, F$_2$, Cl$_2$, Br$_2$. Compounds in which a highly symmetrical solution is likely to survive are underlined.}
        \begin{ruledtabular}
	\begin{tabular}{ccccc}
		X\textbackslash{}Y & O & F & Cl & Br \\ \hline
		P & 0.59 & 0.18 &  \underline{$\approx 0$} & \underline{-12.88} \\ 
		V & 0.34 & 0.65 & \underline{-0.1} & \underline{-7.77} \\ 
		As & 0.078 & 0.04 & 0.25 & \underline{-8.73} \\ 
	\end{tabular}
        \end{ruledtabular}
\end{table}

\begin{figure}
	\centering
	\includegraphics[width=1.0\linewidth]{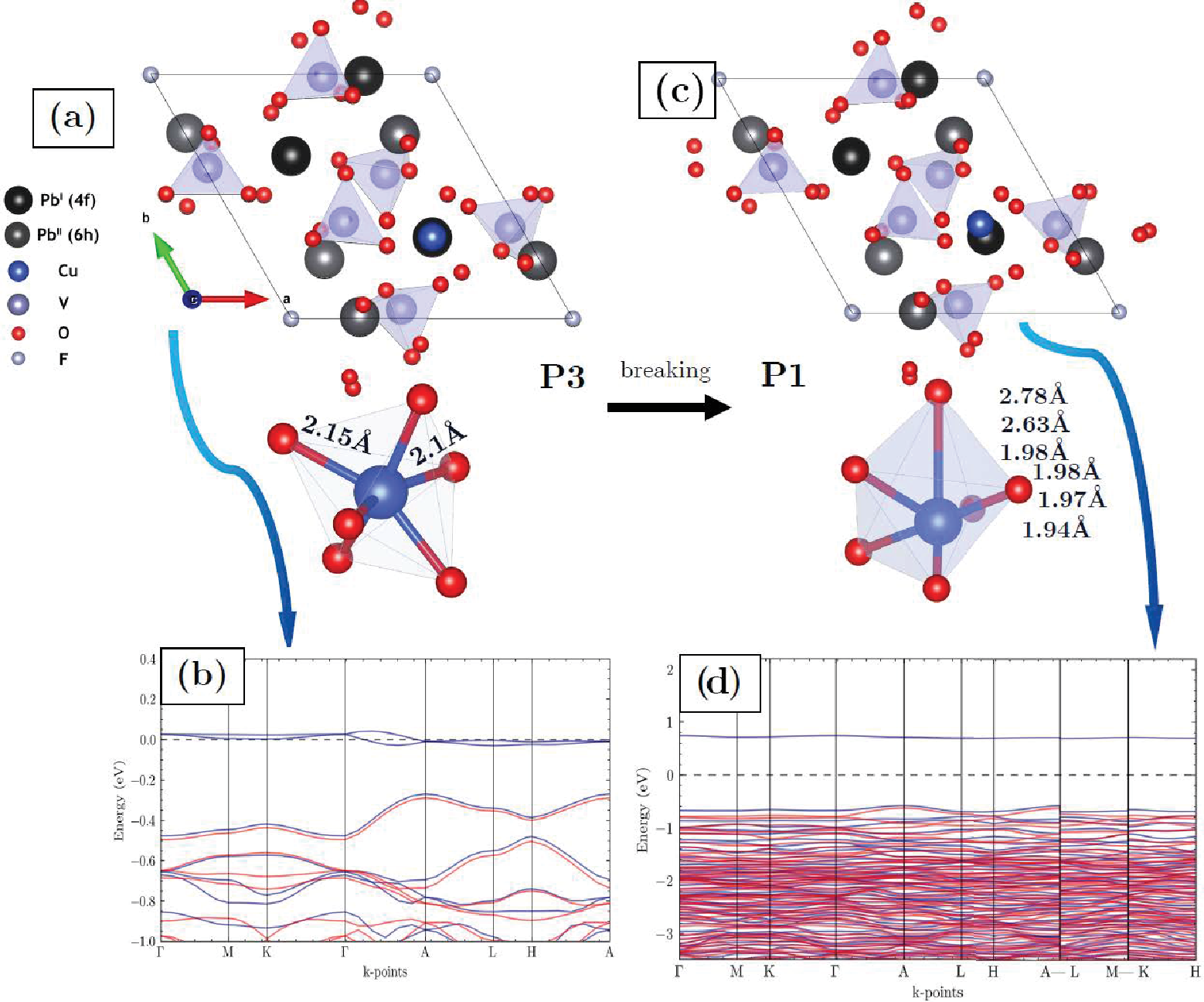} 
	\caption{Effect of crystal symmetry breaking on its electronic structure. (a): Symmetry reduction from $P3$ to $P1$; (b): band structure for trigonal symmetry; (c) top view of the unit cell with broken symmetry; (d) band structure for triclinic symmetry. Compound: Pb$_9$Cu(VO$_4$)$_6$F$_2$}
	\label{fig:sym}
\end{figure}

As seen from the example at Figure \ref{fig:sym}, the breaking of crystalline symmetry, no mater how small the magnitude of atomic displacements, drastically changes the band structure and properties. Once again, full opening of the gap requires breaking of one additional symmetry, and is realised in spin-polarised calculations. This behaviour observed for all considered compounds, with the band structures presented in the Supplementary.

The argument about local $C_3$ symmetry being able to form a complex representation and thus protect the band crossing, also holds here: naturally, $P1$, that remains after the breaking, is unable to do that.

As discussed previously, this crystalline symmetry breaking should not be considered entirely physical. One might imagine performing a powder diffraction measurement of such a sample only to find a hexagonal (trigonal) system, because the local distortions will be averaged out in space and time. However, the resulting gap and semiconducting behaviour are physical; this symmetry breaking is a DFT's way to represent correlation-induced metal-to-insulator transition.

After applying this tree-stage filter - stability, site and symmetry («the ladder») - we see that only two candidates remain - Pb$_9$Cu(VO$_4$)$_6$Br$_2$ and Pb$_9$Cu(VO$_4$)$_6$Cl$_2$. Hence, we shall pay closer attention to the band structure of \underline{Pb$_9$Cu(VO$_4$)$_6$Br$_2$} compound as the most robust choice.

\subsection*{Magnetic order}

\begin{figure}
	\centering
	\includegraphics[width=0.86\linewidth]{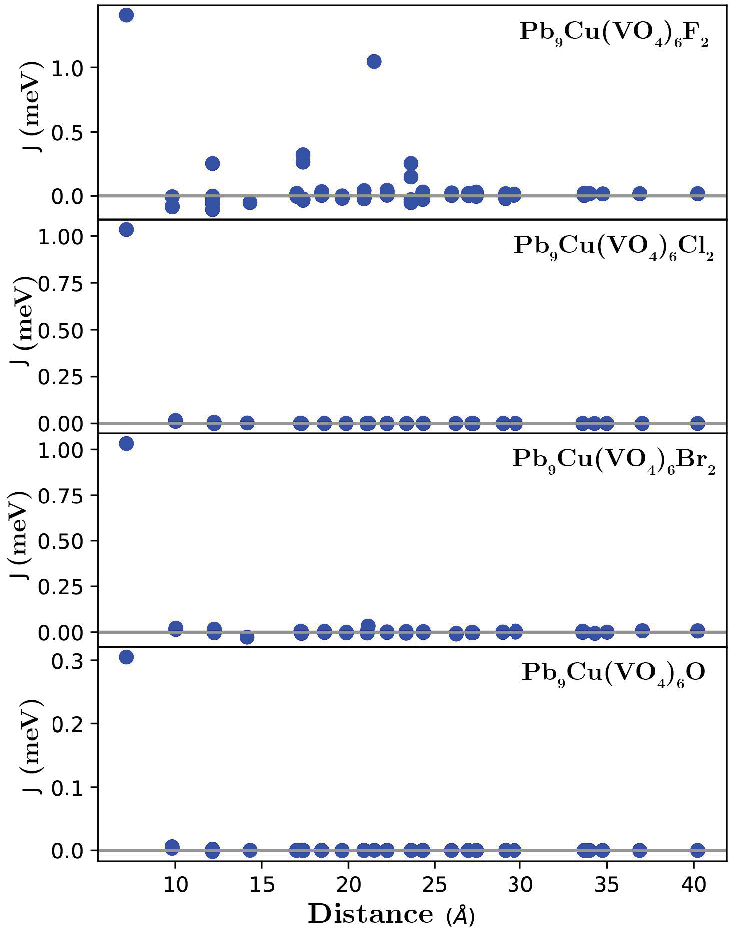} 
	\caption{Dependence of the magnetic exchange parameter $J_{Cu-Cu}$ on the distance between atoms.}
	\label{fig:jvsr}
\end{figure}

Vanadate compounds that satisfied the convex hull criterion were further analysed in regards to their magnetic properties. Heisenberg exchange parameters $J$ were calculated. As a result, we find that the magnetic exchange is dominated by the nearest-neighbour $J_{Cu-Cu}$, which is found to be positive for all studied compounds, namely, Pb$9$Cu(VO$_4$)$_6$Y, Y = F$_2$, Cl$_2$, Br$_2$, O. We may therefore expect the ferromagnetic ground state for them. However, because of the relatively large copper-copper distance of about 7 \AA, $J$ are of the order $\sim 1$ meV, making ferromagnetism unlikely to be observed at room temperature. Magnetic moments and exchange parameters are given in the Table \ref{tab:fm}, and the evolution of $J_{Cu-Cu}$ for vanadate apatites is shown at the Figure \ref{fig:jvsr}.

\begin{table}[b]
	\caption{Parameter $J_{Cu-Cu}$ and magnetic moments of individual atoms for the considered compounds.}
        \begin{ruledtabular}
	\begin{tabular}{ccccc}
		$c$-atom & $J_{NN}$, meV & $m, \mu_B$ & $m_{oxygen}, \mu_B$ & $m_{halogen}, \mu_B$ \\ \hline
		F$_2$ & 1.4 & 0.49 & 0.07 & 0.006 \\ 
		Cl$_2$ & 1.034 & 0.51 & 0.07 & 0.11 \\ 
		Br$_2$ & 1.032 & 0.53 & 0.07 & 0.02 \\ 
		O & 0.305 & 0.56 & 0.06 &  \\ 
	\end{tabular}
        \end{ruledtabular}
	\label{tab:fm}
\end{table}

\subsection*{Band structure}

\begin{figure*}
    \centering
    \includegraphics[width=0.75\textwidth]{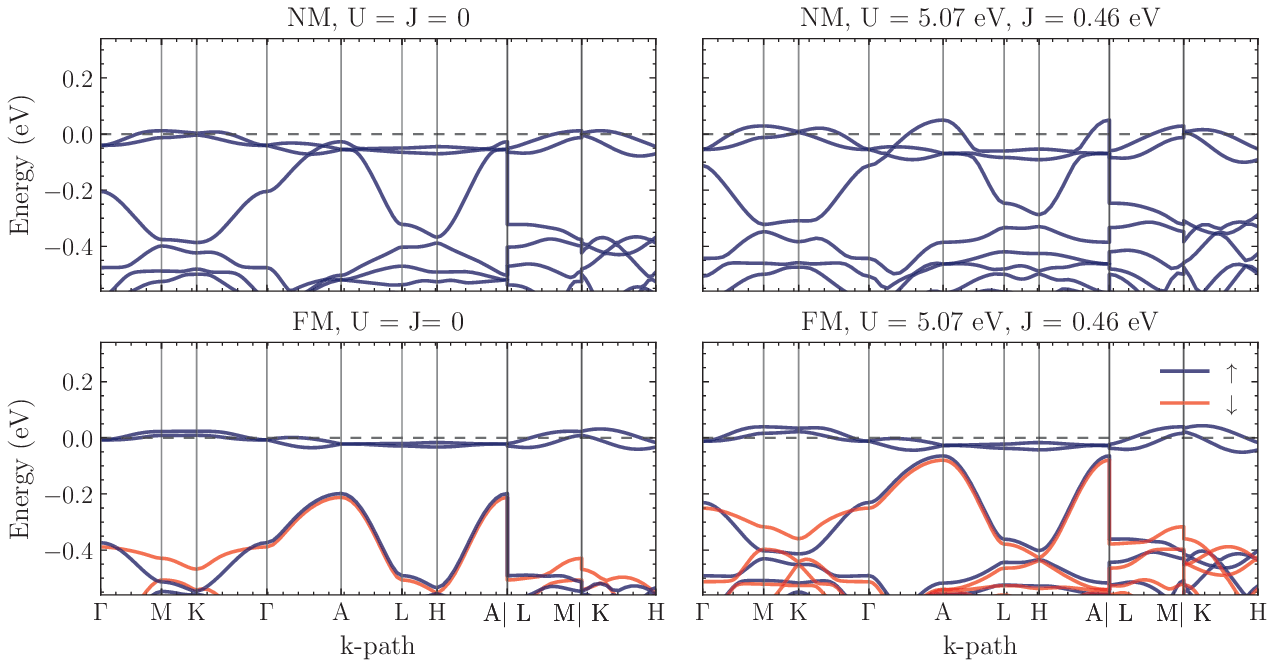} 
    \caption{Near-Fermi band structure of the non-magnetic (NM) and spin-polarised ferromagnetic (FM) Pb$_9$Cu(VO$_4$)$_6$Br$_2$ compound, with and without the application of +U+J corrections.}
    \label{fig:bnd}
\end{figure*}

Figure \ref{fig:bnd} shows the near-Fermi band structure for the non-magnetic and ferromagnetic spin-polarised ferromagnetic cased with and without +U+J parameters. The spin-orbit coupling is neglected here.

At points $\Gamma$ and $A$, the flat bands cross. In the spinless situation, relevant for our non-magnetic variant, the crystal $C_3$ symmetry is accompanied by the time reversal one $\mathcal{T}$ ($\mathcal{T}^{2} = +1$). $\Gamma$ and $A$ can thus be considered as points invariant under the time reversal symmetry (TRIM), and the groups $C_3$ and $\mathcal{T}$ capable of ensuring Weyl-like crossings at $\Gamma$ and $A$, in agreement with \cite{Weyl}. In the collinear ferromagnetic case, $\mathcal{T}$ does not hold, and the symmetry implications require further study, though qualitatively the crossings persist for one of the spin channels.

\begin{figure}
    \centering
    \includegraphics[width=0.95\linewidth]{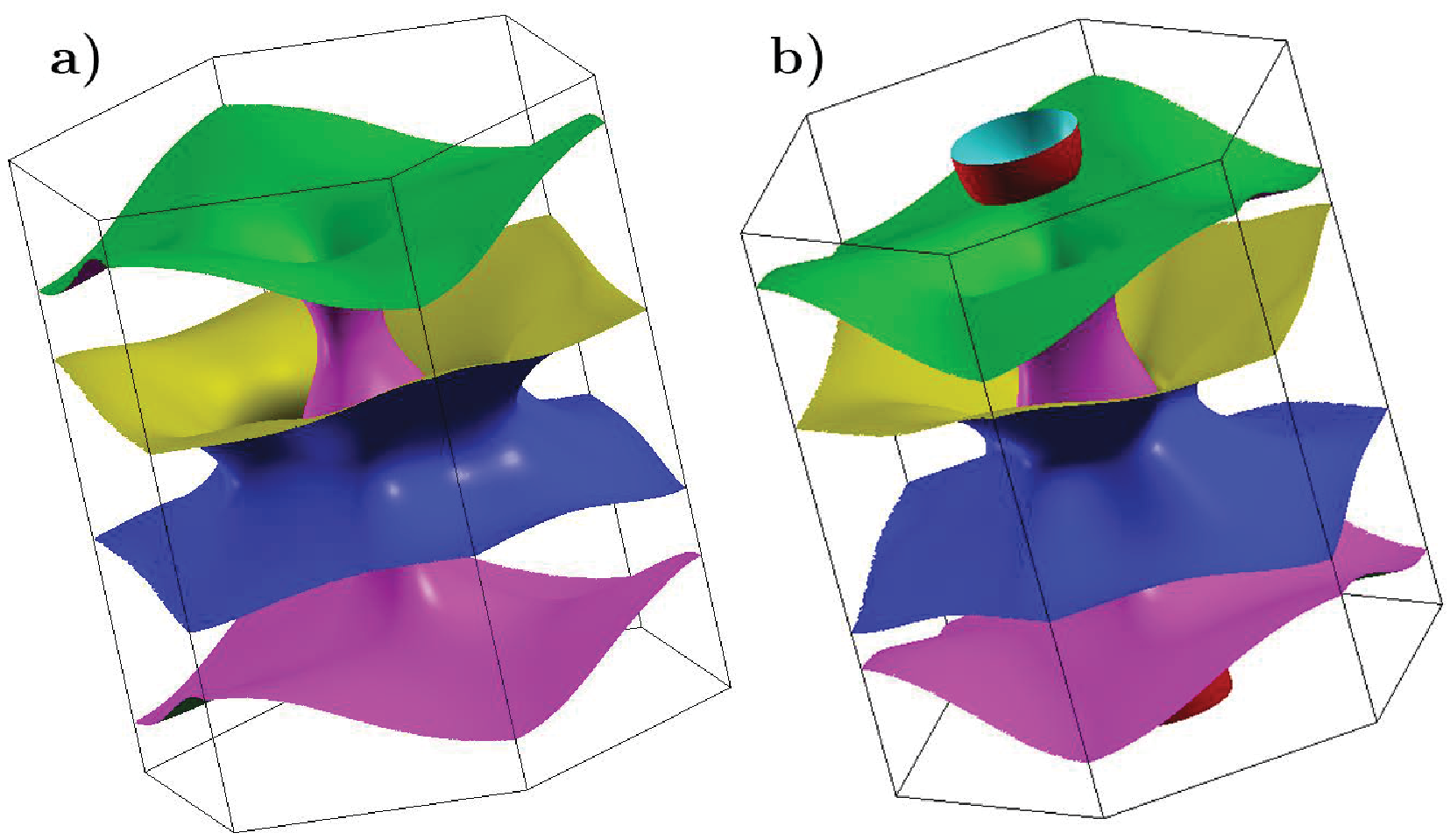}
    \caption{Fermi surface of  Pb$_9$Cu(VO$_4$)$_6$Br$_2$. a) Spin-polarised regime, «spin-up» surface; b) non-magnetic regime.}
    \label{fig:fermi}
\end{figure}

 Figures \ref{fig:bnd}, \ref{fig:fermi} jointly show that the magnetic and non-magnetic band structures of the compound Pb$_9$Cu(VO$_4$)$_6$Br$_2$ have a noticeable difference: in the non-magnetic case, the Fermi level is crossed by an additional band, changing the topology of the Fermi surface.

\begin{figure}
    \centering
    \includegraphics[width=0.95\linewidth]{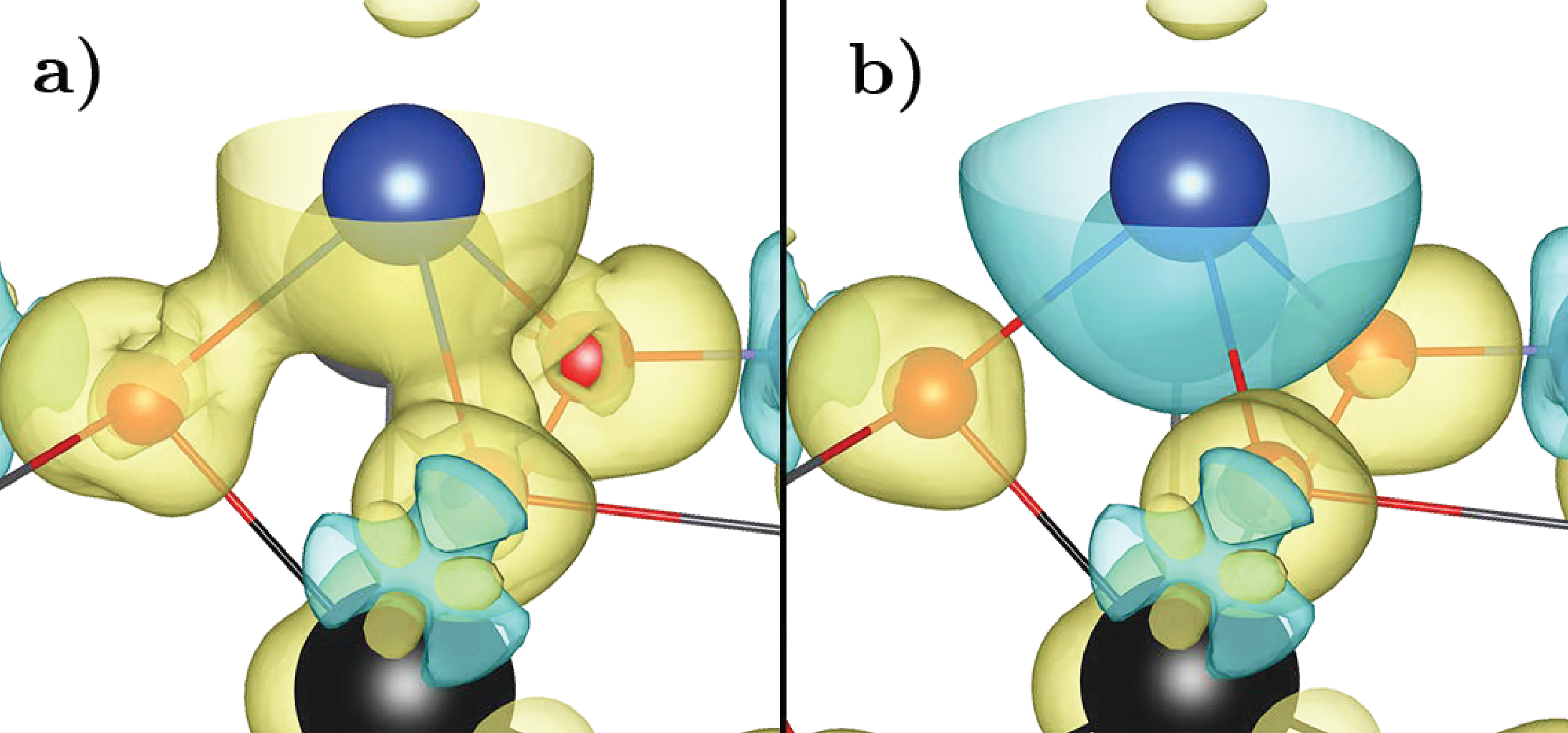}
    \caption{Imbalance of the charge density between spin channels. a) $\uparrow$; b) $\downarrow$.}
    \label{fig:cdd}
\end{figure}

\begin{figure}
    \centering
    \includegraphics[width=0.95\linewidth]{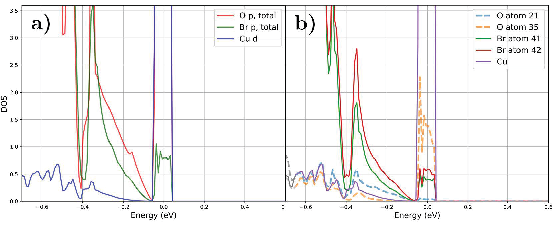}
    \caption{a) Total contribution of atomic species to the near-E$_F$ density of states; b) contribution of individual atoms.}
    \label{fig:cuo}
\end{figure}

\begin{figure}
    \centering
    \includegraphics[width=0.65\linewidth]{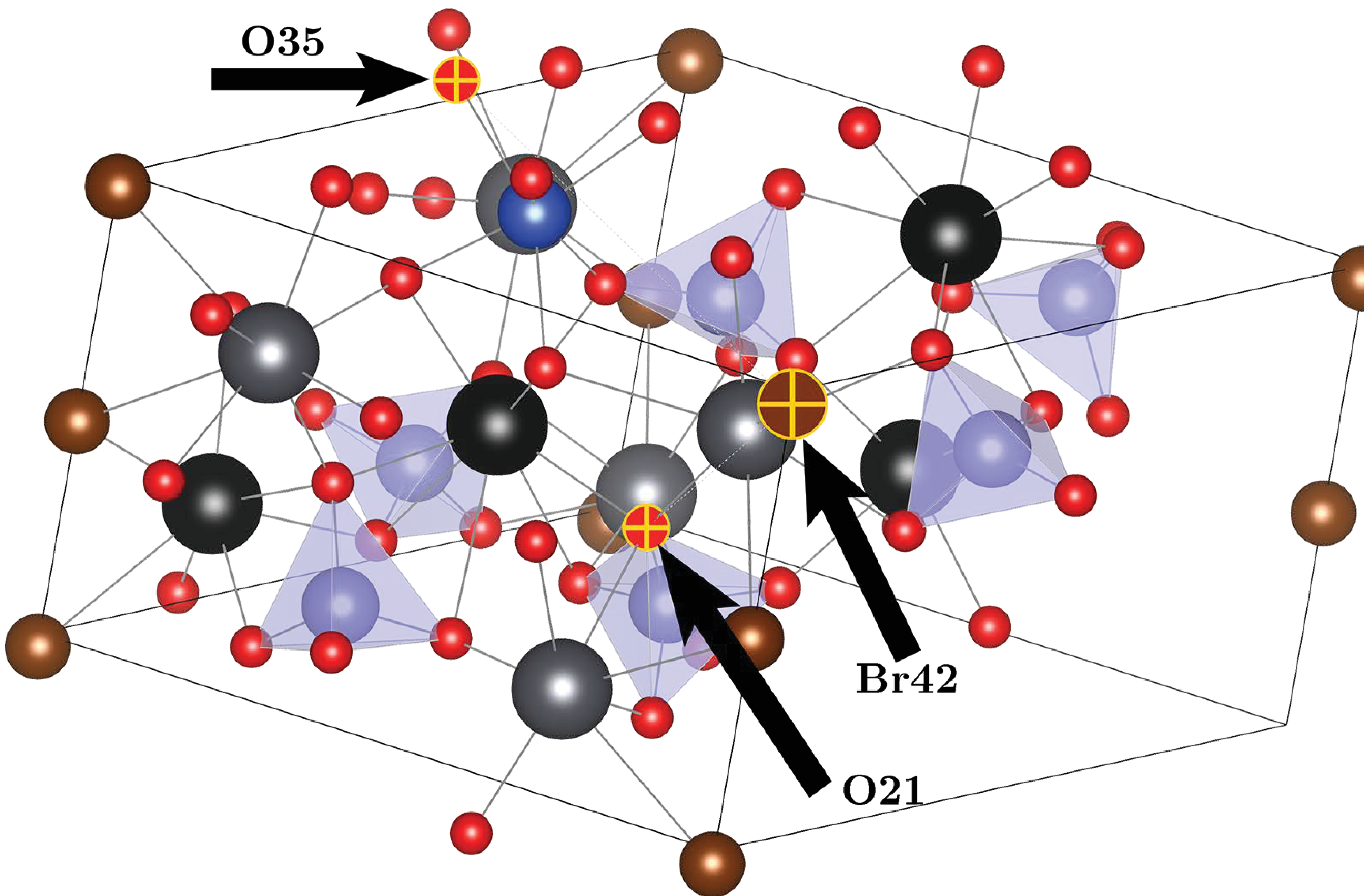}
    \caption{Unit cell of Pb$_9$Cu(VO$_4$)$_6$Br$_2$. Atoms mentioned at Fig. \ref{fig:cuo} are identified.}
    \label{fig:atomsmarded}
\end{figure}

The reason for the lowering of the dispersive $p-$ bands below the Fermi level when spin polarisation is turned on is not symmetry breaking. Instead, we are dealing with a redistribution of charge between the spin channels. When magnetism is turned on, the the total charge is divided between two spin channels - $\uparrow$ and $\downarrow$, and the Hamiltonian is divided into two corresponding blocks. The polarisation of the copper atom means that its charge flows between the channels the most. Partial DOS data shows that the individual contribution to band 3 is highest in the bromine atom (Fig. \ref{fig:cuo}), collectively the weight of the oxygen states in it still prevails, and first of all - oxygen from the Cu-O octahedron. From the perspective of one of such oxygen atoms, the spin-down block senses that its hybridisation partner, Cu $d \downarrow$ , has lost some charge, and thus the hybridisation weakens and it slides down to the lower energy. Since the spin polarisation on oxygen must still remain small, O p $\downarrow$ drags O $p \uparrow$ along with it.

Thus, the hybridisation of the nearest neighbours Cu $d$ - O $p$ becomes \textit{spin-asymmetric}.

\begin{table}[b]
	\caption{The difference between the $\uparrow$- and $\downarrow$- L\"ovdin charges of copper and the oxygen closest to it for the two spin channels, units of $e$.}
        \begin{ruledtabular}
	\begin{tabular}{ccccc}
		& Cu & O & Cu $3d_{xz+yz}$ & O $2p_{x+y}$ \\ \hline
		$q_{\uparrow}- q_{\downarrow}$ & 0.499 & 0.04 & 0.408 & 0.016 \\
	\end{tabular}
        \end{ruledtabular}
        \label{tab:lodwin-spin-pol}
\end{table}

\textit{Conclusions.} We built a simple but revealing screening pipeline: of stability, site preference, and symmetry robustness - and used it to narrow the copper-apatite landscape down to candidates that actually preserve the sought-after near-Fermi band structure portrait. Only the vanadates sit on or below the convex hull, while phosphates and arsenates float above it, making them less promising starting points. 

 Among the two copper sites, substitution at Pb$^{\text{I}}$ ($4f$) consistently leads to the flat band with saddle-point features and symmetry-protected touchings at $\Gamma$ \& $A$ points, whereas  placing Cu at Pb$^{\text{II}}$ ($6h$) instead tends to open a gap or severely distort the bands. The energetic preference for  Pb$^{\text{I}}$ is present but modest (often less than $k_B T$ at 300 K), so some disorder is plausible; yet the electronic argument still points to Pb$^{\text{I}}$ as the relevant configuration. 
 
Finally, symmetry matters: small relaxations that lower $P3$ to $P1$  erase the protected crossings and nudge the system toward a gap. Symmetry-broken DFT here represents the physical outcome usually associated with strong correlations, presenting another appealing case for this approach. 

Applying these three-stage «ladder» leaves a short list, with Pb$_9$Cu(VO$_4$)$_6$Br$_2$ emerging as the most robust candidate for correlated-electron physics. The computed exchange parameter $J_{Cu-Cu}$ is ferromagnetic but tiny (1 meV), making room-temperature order unlikely.

This work narrows the search to a chemically and structurally credible corner where the «flat-band at E$_F$» picture survives first-principles scrutiny, and sets the stage for the future experimental characterisations.

\begin{acknowledgments}
This work was supported by the National Research Foundation of
Ukraine (Grant No. 2023.03/0242).
\end{acknowledgments}




\bibliography{apssamp}

\end{document}